\pdfoutput=1
\documentclass{llncs}
\usepackage{amssymb}
\usepackage{enumitem}
\usepackage{graphicx}
\usepackage[table,xcdraw]{xcolor}
\usepackage{caption}
\usepackage{multirow}
\usepackage{amsmath}
\usepackage{tabu}
\usepackage{wrapfig}
\usepackage{cite}
\usepackage{url}
\usepackage{makeidx}

\usepackage[utf8]{inputenc}
\usepackage[T2A]{fontenc}

\begin{document}

\frontmatter

\pagestyle{headings}

\mainmatter

\title{On Residual CNN in text-dependent speaker verification task}

\author{Egor Malykh\inst{1} \and Sergey Novoselov\inst{1, 2} \and  Oleg Kudashev\inst{1, 2}}

\institute{ITMO University, St.Petersburg, Russia\\
\and
STC-innovations Ltd., St.Petersburg, Russia\\
\email{\{malykh, novoselov, kudashev\}@speechpro.com}}

\titlerunning{Residual CNN in text-dependent speaker verification task}
\authorrunning{Egor Malykh et al.}

\maketitle

\begin{abstract}
Deep learning approaches are still not very common in the speaker verification field. We investigate the possibility of using deep residual convolutional neural network with spectrograms as an input features in the text-dependent speaker verification task. Despite the fact that we were not able to surpass the baseline system in quality, we achieved a quite good results for such a new approach getting an 5.23\% ERR on the RSR2015 evaluation part. Fusion of the baseline and proposed systems outperformed the best individual system by 18\% relatively.

\end{abstract}

\section{Introduction}
I-vector systems are well-known for being state-of-the-art solutions to the text-independent speaker verification task \cite{kenny-fa, kenny, novel, novoselov_non_linear_plda}. Recently, the solution of this task has increasingly been considered from the perspective of deep learning approaches. For instance, ASR deep neural network (DNN) model \cite{novel, novoselov_dnn} divides the acoustic space into senone classes and discriminates the speakers in this space using the classic total variability (TV) model \cite{kenny-fa}. In such phonetic discriminative DNN based systems two main approaches can be distinguished. The first is to use DNN posteriors to calculate Baum-Welch statistics, and the second is to use the bottleneck features in combination with speaker specific features (MFCC) for training the full TV-UBM system. The second approach is considered the most robust to varying conditions \cite{sri_bn}.

As demonstrated by recent publications \cite{themos, larcher, hagay, tdjfa, novoselov_state_plda}, substantial success of the state-of-the-art text-dependent verification systems is mainly due to the progress in text-independent speaker recognition task. Thus, the success of the phonetic discriminative DNN in such a task leads to attempts to use similar approach in text-dependent systems \cite{matejka, zeinali_hmm_ivec, td-prelim}.

In parallel, there are several studies on the use of Deep-Learning approaches aiming to create an end-to-end solutions for discriminating speakers directly in a text-dependent task \cite{microsoft, google}. Such approaches are easily applicable when the duration of the considered utterances is small, since they can be fed as an input of a deep architecture entirely, for example as a spectrogram.

A speaker discriminative approach is the most natural way for speaker verification. \cite{dnnsmall} describes a DNN for extracting a small speaker footprint which can be used to discriminate between speakers.

In this paper we investigate the deep residual CNN \cite{resnet} for direct speaker discrimination. Unlike \cite{microsoft} we focus on the use of spectrograms instead of MFCC as the input features and deep but light residual architecture instead of VGG-like network as the mapping.

\section{Baseline}
A standard i-vector system is used as the baseline in our experiments. The i-vector system models a speech utterance as a low dimensional vector of channel- and speaker-dependent factors using total variability approach, as follows:
\begin{equation*}
s = \mu + Tw,
\end{equation*}
where $s$ is the mean supervector, $\mu$ is the mean supervector of an Universal Background Model (UBM), $T$ is a low rank matrix and $w$ is the i-vector estimated using the Factor Analysis method \cite{kenny-fa}.

We used implementation of the back-end from \cite{zeinali_hmm_ivec}. All i-vectors are length normalized and further regularized using the phrase-dependent  Within-class Covariance Normalization (WCCN). A simple cosine distance scoring is used followed by phrase-dependent s-norm score normalization \cite{tdjfa}.

19 Mel-Frequency Cepstral Coefficients (MFCC) + log energy is used as the baseline features. They are normalized by mean and variance and augmented with $\Delta$ and $\Delta\Delta$. For this system we did not apply voice activity detection.

\section{CNN}

\subsection{Features}
\begin{figure}[h]
  \centering
  \includegraphics[width=0.75 \linewidth]{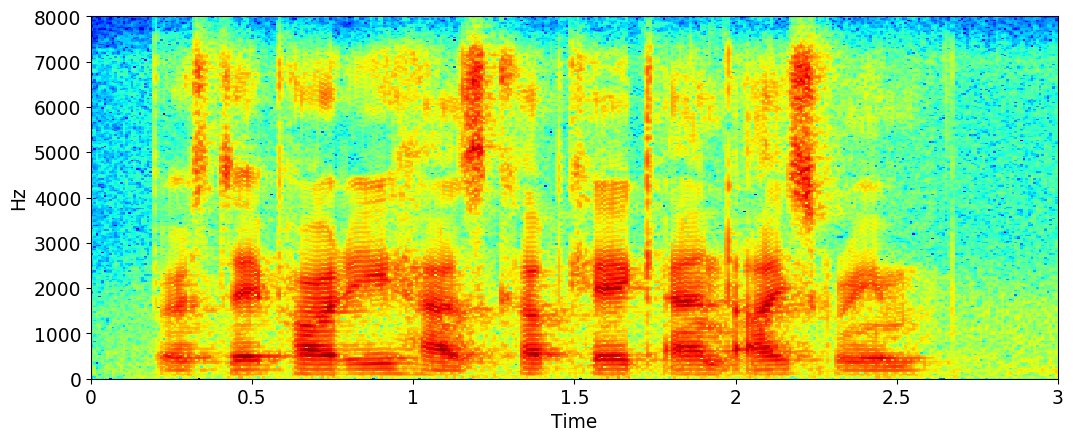}
  \caption{Log power magnitude spectrum of an utterance corresponding to the phrase "Birthday parties have cupcakes and ice cream"}
  \label{fig:spec}
\end{figure}

We use the normalized log power magnitude spectrum obtained via Fast Fourier Transform (FFT) as the input acoustic features for this system. Spectrograms are extracted with the following parameters: window size is $256$, step size is $64$ and Blackman window function is used. Example of such spectrogram is shown in Figure \ref{fig:spec}.

\begin{figure}[h]
  \centering
  \includegraphics[width=0.75 \linewidth]{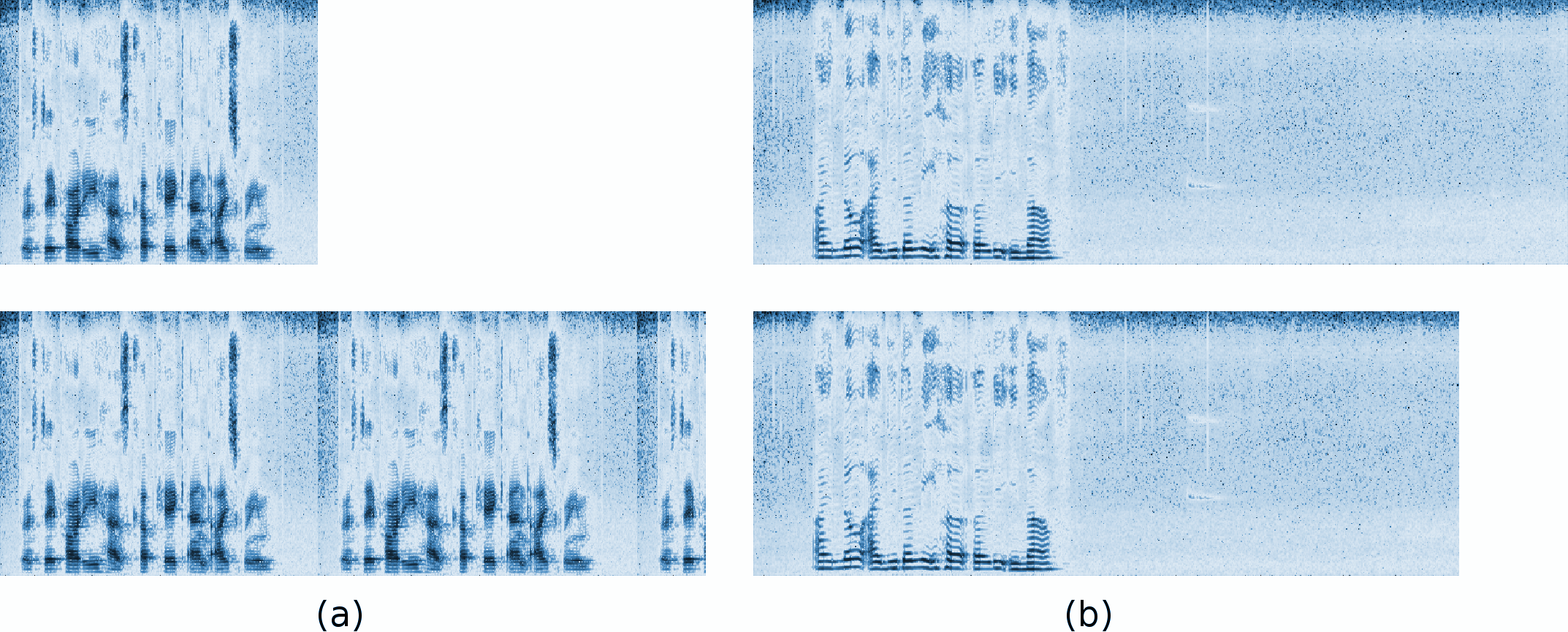}
  \caption{Spectrogram preprocessing for short (a) and long (b) utterances}
  \label{fig:crop}
\end{figure}

The length of the spectrogram along the frequency axis is fixed, but the length along the time axis varies depending on the utterance. However, CNN requires a constant-size image as the input. In order to satisfy this requirement we use the following technique. Images longer than 800 pixels wide are cropped. Images shorter than 800 pixels wide are complimented to the right by their own copy. Such cropping and padding technique is illustrated in Figure \ref{fig:crop}.

\subsection{Residual architecture}
\begin{figure}[h]
  \centering
  \includegraphics[width=0.4 \linewidth]{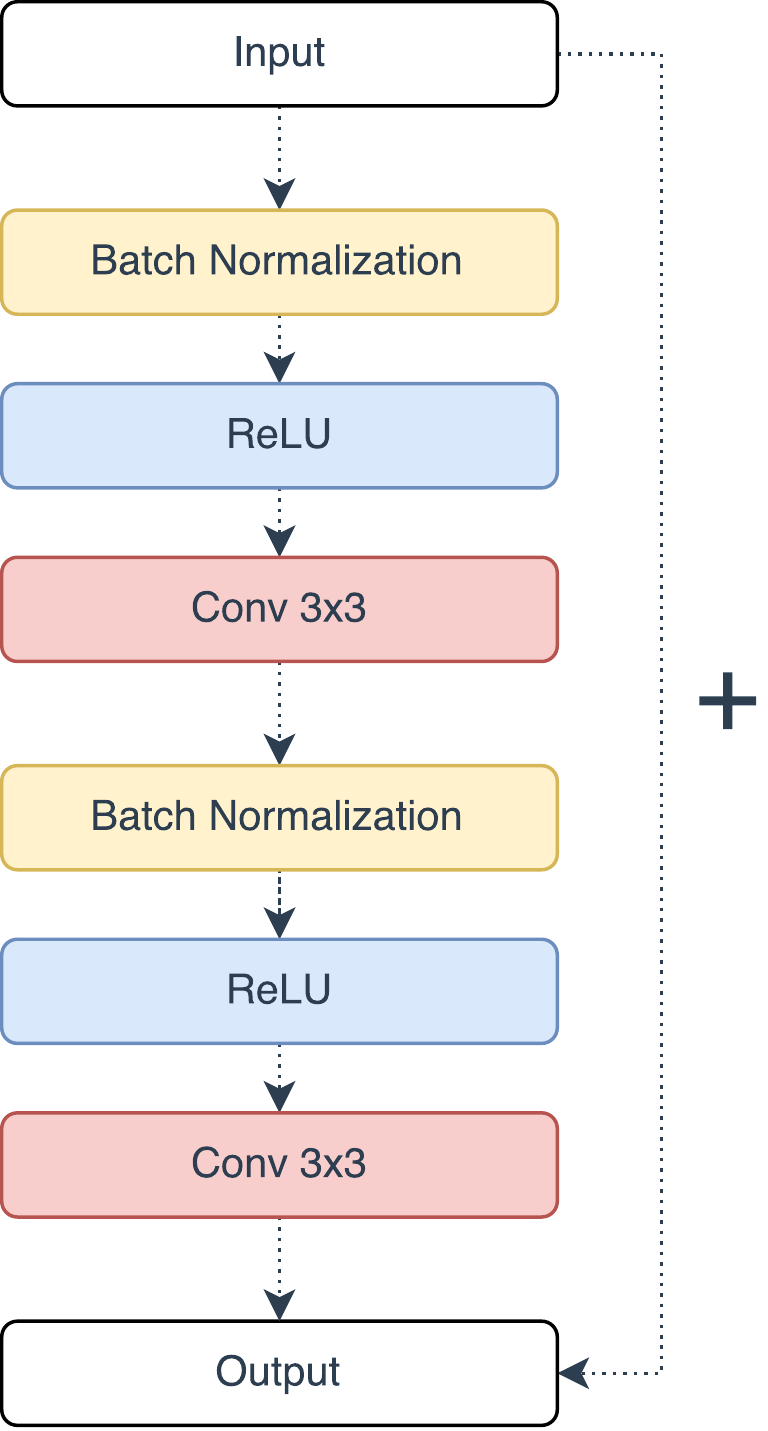}
  \caption{Residual block}
  \label{fig:rb}
\end{figure}

Spectrograms, being two-dimensional tensors, can be considered as images and can be processed by methods used for image processing. Currently, the best convolutional architecture for solving image processing tasks is a Residual CNN \cite{resnet}. Residual architecture is described in \cite{resnet, resnet2} as a stack of several residual units. Residual unit is a mapping
$$x_{l+1} = x_l + \mathcal{F}(x_l, \mathcal{W}_l),$$
where $x_l$ and $x_{l+1}$ are the unit's input and output. $\mathcal{F}$ consists of two $3 \times 3$ convolutions with weights $\mathcal{W}_l$. Additive "shortcut connection" allows the network to satisfy the basic property: adding more layers does not lead to a degradation of the network. Thus, it becomes possible to train very deep networks with a size of 152 or more layers, as shown in the \cite{resnet}. For this study, a network with 18 layers from \cite{resnet} with modifications from \cite{resnet2} was used. Network architecture is shown in table \ref{tab:architecture}. The structure of basic residual block is presented in figure \ref{fig:rb}.

\begin{table}[]
\centering
\caption{Residual CNN architecture}
\label{tab:architecture}
{\tabulinesep=1.2mm
\begin{tabu}{lccr}
layer & kernel/stride & output & \#parameters \\ \hline
Input & $-$ & $257 \times 800 \times 1$ & $0$ \\
Conv+BN+ReLU & $7 \times 7$/$2 \times 2$ & $129 \times 400 \times 64$ & $3.2$K \\
Maximum pooling & $3 \times 3$/$2 \times 2$ & $65 \times 200 \times 64$ & $0$ \\ \hline
Residual block & $3 \times 3$/$1 \times 1$ & $65 \times 200 \times 64$ & $74.1$K \\
 & $3 \times 3$/$1 \times 1$ & & \\
Residual block & $3 \times 3$/$1 \times 1$ & $65 \times 200 \times 64$ & $74.1$K \\
 & $3 \times 3$/$1 \times 1$ & & \\ \hline
Residual block & $3 \times 3$/$2 \times 2$ & $33 \times 100 \times 128$ & $230.1$K \\
 & $3 \times 3$/$1 \times 1$ & & \\
Residual block & $3 \times 3$/$1 \times 1$ & $33 \times 100 \times 128$ & $296.2$K \\
 & $3 \times 3$/$1 \times 1$ & & \\ \hline
Residual block & $3 \times 3$/$2 \times 2$ & $17 \times 50 \times 256$ & $919.8$K \\
 & $3 \times 3$/$1 \times 1$ & & \\
Residual block & $3 \times 3$/$1 \times 1$ & $17 \times 50 \times 256$ & $1\;182.2$K \\
 & $3 \times 3$/$1 \times 1$ & & \\ \hline
Residual block & $3 \times 3$/$2 \times 2$ & $9 \times 25 \times 512$ & $3\;674.7$K \\
 & $3 \times 3$/$1 \times 1$ & & \\
Residual block & $3 \times 3$/$1 \times 1$ & $9 \times 25 \times 512$ & $4\;723.7$K \\
 & $3 \times 3$/$1 \times 1$ & & \\ \hline
Average pooling & $-$ & $512$ & $0$ \\ \hline
SoftMax & $-$ & $97$ & $50$K \\ \hline
\textbf{Total} & & & $11\;228.0$K
\end{tabu}}
\end{table}

\section{Experimental setup}

\subsection{RSR2015 corpus}
In our experiments we use the RSR2015 database \cite{rsr2015}. The RSR2015 provides data for three main use-case verification scenarios:

\begin{itemize}

    \item \textbf{unique pass-phrase}: each client pronounces the
same pass-phrase,

    \item \textbf{user-dependent pass-phrase}: each client
pronounces his or her own pass-phrase,

    \item \textbf{prompted text}: each client pronounces a sentence prompted by the system.

\end{itemize}

In this paper, our focus is on the first use-case where each speaker pronounces a particular sentence. The RSR2015 database contains audio recordings from 300 speakers (143 female and 157 male). There are 9 sessions for each of the participants. Each session consists of 30 short sentences. The database is collected in the office environment using six different portable recording devices (four smartphones and two tablets). Each speaker was recorded using three random different devices out of the six.

The database is randomly split into three non-overlapping groups of speakers, one for background training, one for development stage and one for evaluation stage. The number of male/female speakers is balanced for each group: 50/47 in the background set, 50/47 in the development set and 57/49 in the evaluation set.

We use the background set only for training our speaker verification systems. The development set is used to estimate calibration and fusion parameters. All test trials are performed on the evaluation set.

We focuse only on the scenario where the speaker pronounces correct pass-phrase. All experiments are conducted according to the part 1 protocols of the RSR2015 database. We consider pooled male and female trials for system performance measure.

Extended training set which contains the background and development sets is used in additional experiment.

\subsection{Baseline}
Parameters of WCCN matrix and i-vector extractor are estimated using the background subset of the RSR2015 corpus only. As described in \cite{zeinali_hmm_ivec}, we use the following representation of the WCCN matrix:
$$\overline{W} = W + \frac{1}{2} E,$$
where $E$ is the unit matrix of appropriate dimensionality. This trick helps to prevent an overfitting despite the small number of speakers in the background subset.

\subsection{CNN}
CNN is implemented using the Keras framework \cite{keras} on top of the TensorFlow \cite{tensorflow} backend. ADAM optimizer \cite{adam} with learning rate set at $10^{-4}$ is used for training

Network is trained to discriminate between all speakers in training set using the softmax layer and categorical cross-entropy loss function. In the evaluation phase an output from the 512-dimensional (same as i-vector) penultimate layer is used as the embedding corresponding to the input utterance.

\section{Results and discussion}

\begin{figure}[h]
  \centering
  \includegraphics[width=0.75\linewidth]{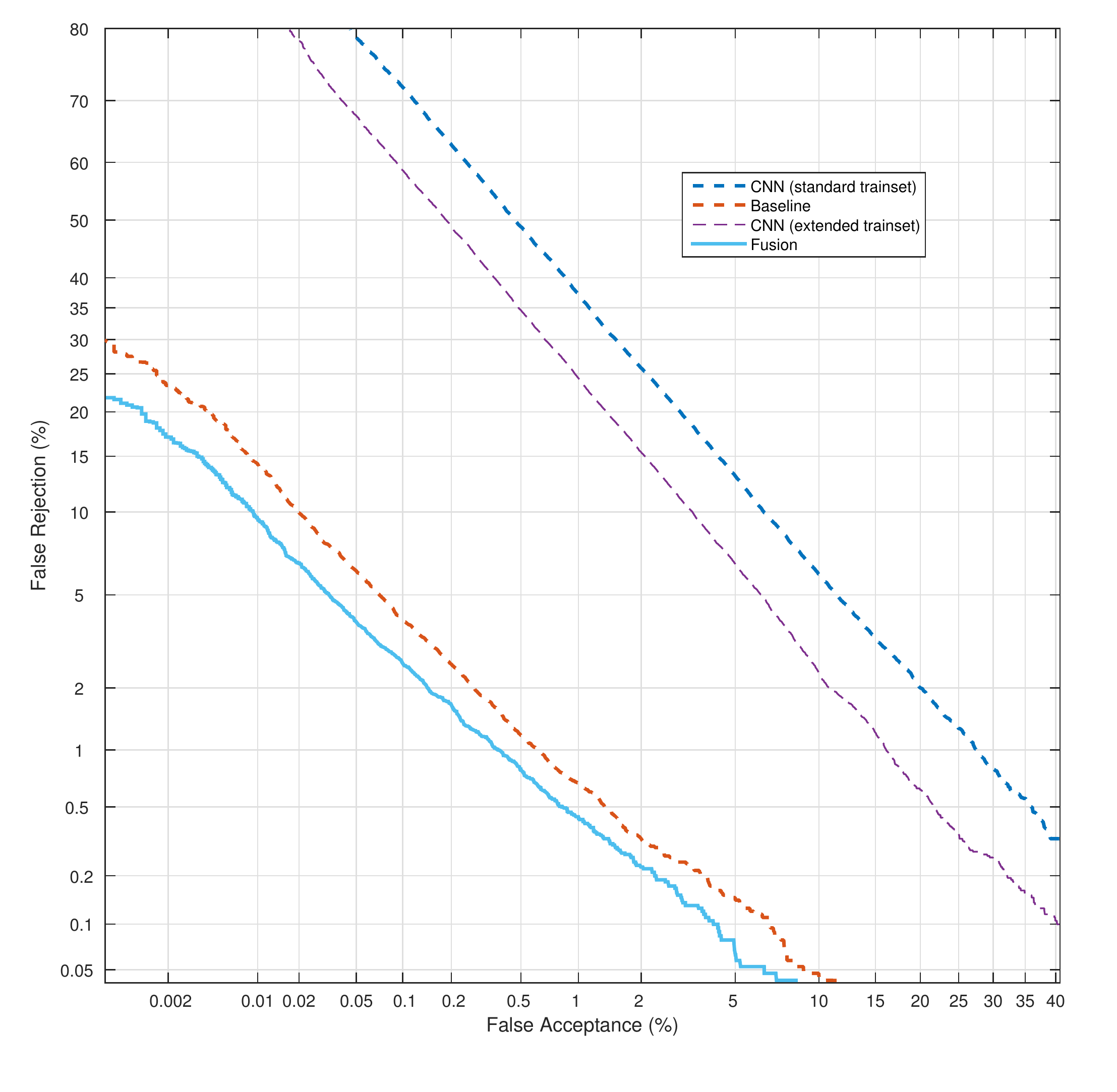}
  \caption{DET curves for the RSR2015 evaluation part}
  \label{fig:det}
\end{figure}

The result of our research is presented in Table \ref{tab:results} in terms of the Equal Error Rate (EER) and the minimum detection cost function (minDCF) with $P_{\textrm{tar}} = 10^{-3}$. Baseline system demonstrated a very good result with an EER of less than 1\% which is comparable with the result from \cite{zeinali_hmm_ivec}. Deep CNN system achieved an EER of 6.02\%. Fusion of this two systems shows 18\% relative improvement over the baseline system which is the evidence of the fact that classic i-vector systems and deep learning systems results in decorrelated embeddings and thus can be used together.

Relatively poor performance of the system under investigation can be explained by the small size of the training set (97 speakers). Such conditions leads to overfitting of discriminative model. The hypothesis is that the deep residual CNN requires much more data for training and expanding training set will lead to a significant increase in accuracy. Experiments on the extended training set (194 speakers) sustains it resulting in an 5.23\% EER. We hope that deep learning approaches will be able to outperform the i-vector based systems in the future.

\begin{figure}[h]
  \centering
  \includegraphics[width=0.75\linewidth]{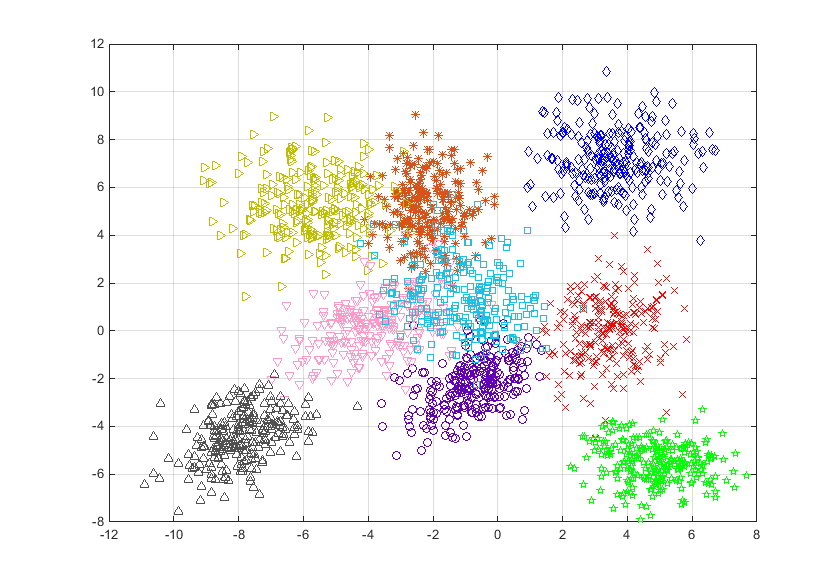}
  \caption{Projection of embeddings to two main principal axis for 9 speakers}
  \label{fig:proj}
\end{figure}

Figure \ref{fig:proj} illustrates the projection of CNN embeddings of the 9 randomly chosen speakers on two principal axis using the Principal Component Analysis. DET-curves of the all considered methods are shown in Figure \ref{fig:det}.

\begin{table}[h]
    \caption{Evaluation results in terms of EER [\%] and minDCF}
    \label{tab:results}
    \begin{center}
    \begin{tabular}{lr@{\quad}r}
    \hline\rule{0pt}{12pt}
    \textbf{System} & \textbf{EER} & \textbf{minDCF}\\[2pt]
    \hline\rule{0pt}{12pt}
    Baseline  & $0.79$ & $0.23$\\
    Deep CNN  & $6.02$ & $0.94$\\
    Deep CNN (ext) & $5.23$ & $0.92$\\
    Fusion    & $\textbf{0.64}$ & $\textbf{0.18}$\\
    \hline
    \end{tabular}
    \end{center}
\end{table}

\section{Conclusion}
In this paper, we presented studies of deep residual CNN architecture in the task of text-dependent verification. Raw normalized spectrograms of speech signals is used as the input features. Experiments conducted on Part 1 of the RSR2015 database showed that despite the small amount of training data, it is possible to train a deep speaker embeddings extractor, which makes it possible to separate the speaker classes fairly well. Best achieved result of the individual system is an 5.23\% EER.

We also showed that increasing the amount of training data leads to the expected strengthening of the extractor and improvement of the results. Our future work will be focused on the improving the quality of deep CNN based systems and bringing them to the level of baseline i-vector systems. It can be noted already that fusion of the deep CNN and i-vector extractors gives a good performance gain of 18\% relative improvement.

\section{Acknowledgements}

This work was financially supported by the Ministry of Education and Science of the Russian Federation, contract 14.578.21.0126 (ID RFMEFI57815X0126).


\end{document}